# Evolving inductive generalization via genetic self-assembly

Rudolf M. Füchslin*, Thomas Maeke, Uwe Tangen & John S. McCaskill

Ruhr-Universität Bochum, Biomolecular Information Processing (BioMIP),

Schloss Birlinghoven, D-53754 Sankt Augustin, Germany.

*Corresponding author: rudolf.fuechslin@biomip.ruhr-uni-bochum.de

## Abstract

We propose that genetic encoding of self-assembling components greatly enhances the evolution of complex systems and provides an efficient platform for inductive generalization, i.e. the inductive derivation of a solution to a problem with a potentially infinite number of instances from a limited set of test examples. We exemplify this in simulations by evolving scalable circuitry for several problems. One of them, digital multiplication, has been intensively studied in recent years, where hitherto the evolutionary design of only specific small multipliers was achieved. The fact that this and other problems can be solved in full generality employing self-assembly sheds light on the evolutionary role of self-assembly in biology and is of relevance for the design of complex systems in nano- and bionanotechnology.

# 1 Introduction

Understanding the autonomous design of self-assembling complex systems[1] is vital for further developments in nanoscience. Despite much progress in harnessing evolutionary processes[2,3] and in particular genetic algorithms[4,5], the conditions for evolving general solutions to problems, applicable to a combinatorially complex variety of distinct problem instances, remain unresolved.

We are specifically interested in those cases where this variety prohibits an evolutionary construction of a general solution by accumulating specific solutions to individual problem instances. In such a situation, a general solution has to be constructed by evolutionarily detecting and exploiting the underlying structural properties of the whole task under consideration.

Such general solutions are economical and useful as modules for building complex systems. Natural instances abound at the nanoscale: for example general sequence replication solved by polymerases and base pairing, general protein biosynthesis solved by translation with the ribosomal apparatus (for the evolution of the genetic code see[6,7]), and general pathogen recognition[8] solved by mRNA splicing between sets of sequence modules to create antibody diversity.

In this article, we show by explicit simulation that complex circuit design problems can be solved by exploiting the properties of self-assembling, genetically encoded components. We posit that natural systems have evolved general solutions to environmental tasks efficiently by making use of a similar modular genetic encoding of self-assembling units.

To elucidate the evolutionary capabilities of self-assembling systems, we discuss in detail the evolution of scalable digital multipliers. The multiplication problem can serve as a prototype of a complex convolution in the genotype-phenotype mapping because the interior bits of multiplication products are notoriously convoluted functions of the inputs. They even find application in random number generation[9,10]. Scalability in this context means that by employing one and the same set of self-assembling components, arbitrarily large *n x n-bit* functional circuits will be constructed given sufficient resources of space and numbers of component copies. Full scalability necessarily implies that the components, by virtue of their internal logic and the patterns they form by self-assembly, embody the logical structure of multiplication in abstract generality.

The methods presented are not restricted to multiplication. Other scalable circuit designs, such as general arithmetic logic units (ALU), adders or a binary to Gray code converter have also been successfully evolved.

Self-assembly is a process by which the local interaction between components (e.g. based on shape complementarity), determines their assembly into larger structures[11] and is vital for biological systems. It is a structuring process complementing catalytic rate control and still operating near equilibrium, which, when genetically tuned, allows macroscopic objects to be constructed reliably under varying conditions. The diversity and precision of nanoscale biological function leads one to expect that complex engineering structures such as nanoscale circuits[12,13,14] may also be assembled from components equipped with analogous recognition elements. Self-assembly is also important in the design of self-replicating molecules[15,16], supramolecular chemistry[17], natural and artificial cells[11,] and in molecular computation[18,19,20]. Fuelled by progress at the micro-scale[21,22,23], the use of self-assembling nanostructures holds the promise of surmounting the physical limits to lithographic instruction[24,25]. Microscopic planar self-assembly of electronic components, with subsequent regular wiring completion (e.g. by electroplating) has been demonstrated in the laboratory[26], and this would also allow physical wiring completion of the self-assembling functional architecture investigated here. Recently, also three-dimensional mulitcomponent self-assembly of electronic components has been shown[27]. These developments argue that the evolution of self-assembling components can have an immediate impact on nanotechnology.

Physical models for evolution involving self-assembly either explicitly[28] or implicitly as in the quasispecies theory[29], have only addressed the evolution of specific solutions to survival problems, and hence have not revealed an evolution-enabling role for general problem solution via self-assembly. Indeed, in the case of complex systems, autonomous evolutionary design has foundered for complex problems on three features of the evolutionary process: the ruggedness and the sparseness of good solutions in the genotype to phenotype mapping[30], the small fraction of the physical environment that an individual experiences[31] and non-monotonic optimization as a result of frequency dependent selection.

The current work points to potential advantages in biasing evolutionary search to favor general solutions for complex problems. Practically, *de novo* circuit design has proved a formidable barrier for artificial evolution for two reasons. First it has been practically impossible to evolve all but the simplest digital circuits using examples of correct behavior, furthermore structures evolved in this way proved to be idiosyncratic and irregular, difficult to use in a modular way or to generalize. Self-assembling components overcome both of these difficulties.

Larger digital multiplier circuits, constructed using information on only the correct output from limited multiplication examples, have been beyond the limits of evolutionary design. Even rationally designing a multiplier circuit from primitive components is provably hard for minimal resources'. Unbiased genetic algorithms[32,33,34,35] have found only special circuits for multiplying very small numbers, and searching general feed-forward circuits to find even a non-minimal multiplier has proved tractable for similarly sized small problems[34,35]. In fact, the largest binary multiplier circuit found with unconstrained search' achieved the multiplication only of 4-bit numbers and employed special logic that does not generalize to larger numbers.

We report that scalable circuits can be designed when self-assembling genetic units are introduced into the evolutionary design process. Fig. 1, referring to a general multiplier, gives an overview of this process that will be detailed further in the following. The figure shows self-assembling logic blocks (SLBs), to be discussed in detail in Sec. 2.1. A genome encodes the computational logic and the recognition sites of a limited number (typically six to ten) of different types of SLBs. These SLBs spontaneously aggregate to form complete logic circuits whereby it is assumed that sufficient identical copies of each SLB are available. In this work, the recognition sites determine the docking of components to form a two-dimensional electrically connected array. This is achieved in simulation by providing (virtual) quadratic substrate boards onto which the SLBs assemble. The self-assembly principle remains valid in free solution, so the board is not a necessary feature of the presented method. The self-assembly process and the details of the recognition mechanism or the initialization of the substrate are presented in Sec. 2.2.

We emphasize that the evolved circuits are scalable: the local matching rules allow multipliers of any desired overall size to be assembled by simply using more component copies (and correspondingly large substrate boards). The intrinsic possibility for a (potentially complex) global geometric regularity of self-assembled circuits is the basis for the evolution of scalable solutions to problems which are logically complicated but feature abstract internal regularity.

In an autonomous evolutionary design process, one has to compare circuits representing solutions (or partial solutions) by referring exclusively to their outputs and not by externally qualified internal structures. Additionally, the fitness function has to be chosen to meet the requirements of scalable designs, see Sec. 2.3. One has to consider that evolutionary progress can result from two basic effects: structural improvements that lead to an enhanced performance on all (or a subclass of all) possible problem instances or erratic improvements resulting from adaptation to specific problem instances. In the case of multiplication, an example for the former is a circuit that realizes multiplications by powers of two via bit-shifts (note that in binary notation a multiplication by a power of two is only a bit shift) and therefore is able to deal with all possible multiplications of the form *a*$*2^n$. On the other hand, an erratic improvement would result from accidentally acquiring the ability to reproduce the correct result of say *37\*16* without an increase in performance on other multiplications. For the evolution of scalable solutions, only structural improvements are desired: erratic progress is not only nugatory, but may even lead a population into a hard-to-escape local maximum of fitness. Scalable circuits are necessarily based on structural improvements; adaptation to specific problem instances without exploiting underlying logical structures is intrinsically limited, simply because the genome encoding the circuit components is of finite size.

In order to exploit the intrinsic possibility for structural improvements resulting from employing self-assembling designs, we set up an autonomously regulated evolution scheme. This is achieved in the following way: besides the logical functions and recognition sites for a set of SLBs, each individual genome encodes a small evolvable set of problem instances (the "test vector"). The algorithm exploits frequency dependent selection: When individuals contest with respect to multiplication ability, one individual's circuit is scored on its opponent's test vector in pair-wise tournament selection, see Fig. 1e and Sec. 2.4. This differs from twin population co-evolutionary optimization[36], in that offspring have to cope with the test vectors of their siblings. Even small test vectors (e.g. size 16) proved to be sufficient for the evolution of arbitrarily large multipliers or ALUs. The genetically linked "co-evolution" of test-vectors and circuit designs drove the population of test vectors automatically at a manageable rate towards the most convoluted multiplication tasks.

## 2   Methods

Genetic self-assembly involves four aspects for which we detail our method below:

1. The self-assembling logic block (SLB) and its corresponding gene, encoding both, inter-block recognition sites and logical functionality.

2. The self-assembly and circuit synthesis.

3. The evaluation of these circuits using test vectors (lists of problem instances, here either a single number or a pair of numbers, as e.g. in the case of multiplication). A fitness function is provided that yields a modular quantification of partial success; this is a requirement for evolving scalable circuitry.

4. The evolutionary dynamics of populations of interacting proliferating individuals, including structured mechanisms of variation for encoded SLBs and test vectors.

The interplay of encoding, self-assembly and evolutionary dynamics is shown in Fig. 2. Note that the genome of an individual carries information for three different types of entities used at distinct stages of the evaluation process: recognition patterns determining the self-assembly process, logic structures defining the functionality of the SLBs and finally a test vector, used in a tournament selection process. The fact that these different entities are encoded on the same genome leads to a coupling of their evolution.

The specific model choices and parameters we discuss in the following seem to be the most natural and were chosen on the basis of simplicity, but some of them did prove critical for achieving rapid evolutionary optimization. In order to stress the distinction between basic properties of self-assembly and specific technical model choices, several of the latter are discussed in the appendix. This split also emphasizes the fact that the more general aspects of self-assembly are of fundamental relevance for successful evolution, whereas most of the technical conventions proved to be convenient or beneficial with respect to efficient evolution but not critical for success as such. In consequence, Fig. 3 to 5 refer jointly to Sec. 2 and the appendix. A complete list of all parameters and variables is given in Table 1.

## 2.1 Self Assembling Logic Blocks (SLB)

The structure of a SLB is shown in Fig. 3a. The computational functionality is determined by four outputs (o0 to o3), each of which gives a signal that is a function of the four input signals (i0-i3). It would be possible to calculate each of the outputs using a four-bit function generator, making simple signal transfer (an output is directly connected to an input) rather sparse in the space of genotypes. This difficulty can be overcome by an encoding representing a phenotypical function as given in Fig. 3b, which leads to a natural bias towards signal transfer. The details of this encoding are not critical, only the fact that it establishes a balance between function and routing. For the implementation chosen in this work see the appendix.

The heterophilic recognition sites, represented in Fig. 3a by sequences of sockets (on the left and the upper edge) and plugs (on the right and the lower edge), are the basic structures determining the self-assembly process of the SLBs into a rectangular array. Our investigations showed that the recognition mechanism should exhibit the following features (for our implementation see the appendix). First, the balance between variability of self-assembly patterns and evolutionary efficiency is critical. In our simulations, this balance is controlled by the length of the recognition sequence and the size of the alphabet in use. And second, an evolutionary freedom to make positions in the recognition sequence promiscuous (in our implementation by the possibility to equip a position in the recognition sequence with no plug or socket). From an abstract point of view, this means that the evolution of a pattern can be achieved employing two different mechanisms, possibly in combination. Firstly, defining a pattern by constructing according recognition sequences and secondly, establishing a pattern starting from promiscuity by progressive exclusion of matches between specific types of components.

The genome usually contains information for four to ten different SLBs with fully evolvable logic and recognition sites (each requiring 96 bits, see the appendix). Additionally, it encodes one auxiliary default block, which has only evolvable plugs and a fixed simplest logical functionality, namely just transmitting inputs to outputs, see Fig. 3d. Note that due to the fact that the socket recognition sites of this default block remain empty by definition over the whole course of evolution, this special SLB matches any combination of plugs and therefore ensures complete self-assembly for any genome. In all the results presented, the logical functionality of this default block is not evolvable. However, this turned out not to be critical.

The encoded SLBs may differ both in their internal logic functions and in their recognition patterns. In order to restrict attention initially to feed-forward circuits, we consider logical components with only inputs on two edges (top and left) and outputs on the other two edges. Thus, the square tiles are not invested with rotational degrees of freedom in this simple case. Hexagonal or other shaped tiles could also have been chosen. Having just two connections per edge was found to provide a suitable granularity for assembling complex digital processing. A simpler structure with one input or output per edge only allows two 2-input combinatorial functions per SLB and, while many such blocks can emulate the functionality of SLBs with two connections per edge, this does not provide a good balance between routing resources and logic. It turned out that restricting the maximal amount of logic in the SLBs may speed up evolution. This was implemented by requiring that only a given number $n_{FG} < 4$ of the outputs of an SLB delivers a signal from a four-bit function generator, whereas the remaining outputs are connected either to ground or directly to an input. In the case of multiplication, $n_{FG} < 2$ proved beneficial (although not crucial), whereas ALUs could only be evolved by allowing maximal use of function generators $n_{FG} = 4$. The generalization of the presented structural elements and mechanisms to three dimensions is straightforward.

## 2.2 Self-assembly process and circuit synthesis

Both, the logical interconnect and the overall logical circuit, are specified uniquely by the block self-assembly, mediated by the recognition/binding mechanism and taking place on a quadratic board (serving as a substrate) that determines the overall size of the circuit and provides the interface to the environment. Fig. 4 shows a board for only a small circuit, suitable for *2×2-bit*-functions. Such a function *f(x,y)→z* has a 4bit output, denoted by $Z_0$-$Z_3$. In order to allow simplify routing, the input signals $X_0$-$X_1$ and $Y_0$-$Y_1$ are provided redundantly and remaining open inputs are connected to ground. Note that for $n_{bit} \times n_{bit}$-problems we employ a board of size *$4n_{bit} \times 4n_{bit}$* as required by the presented interface to the input signals.

For initial concreteness and computation speed, the self-assembly process is chosen to be completely deterministic, whereby a SLB can be appended if the left and upper plugs do not mismatch (see above and Fig. 1c and 4). To initiate the self-assembly, the board provides two outer rims with evolvable, repetitive binding sites, acting as initialization for the self-assembly of the SLBs, which themselves are assumed to be available in as many copies as are necessary to complete the circuit.

The recognition sites, as they are given in a simplified form in Fig. 4, lead to unequivocal matching of the SLBs but this is not the generic case. In order to resolve ambiguities, a binding energy is employed, i.e. if different SLBs match with a given plug-structure, the one leading to the higher binding energy is taken. The binding energy or binding quality is given by the number of truly matching plug-socket pairs, whereby promiscuous matches are not counted. If there still remains an ambiguity, the SLB encoded at the largest distance from the start of the genome is taken. If no fully evolvable SLB matches, the genome's default SLB is plugged in; this is always possible due to the fact that its socket recognition region is fully promiscuous.

Up to this point, the outer rims of the board have no functionality and their evolvability is restricted to a single repeated recognition site. A straightforward generalization is given by encoding a finite number of additional edge blocks on the genome and allowing edge self-assembly, illustrated in Fig. 5 and detailed in the appendix. Self-assembly of the edges proved beneficial for flexible evolution and was necessary e.g. in case of ALUs. In addition to establishing more complex recognition patterns on the rims of the board, the interface to the input signals is extended by edge blocks carrying some (simple) evolvable functionality. Based on the observation that the input structure and its evolvability is relevant for evolution speed, we plan in future to allow the system to assemble its own inputs and outputs at arbitrary positions, *via* the generic inter-component recognition mechanism, treating inputs and outputs as blocks with their own recognition patterns.

Examples for a scalable multiplier and an ALU are given in Fig. 6. Fig. 6a represents the different types of SLBs employed for the circuit, whereas Fig. 6c shows the self-assembly process. The scalability of the circuits was shown by identifying the detailed logical functionality of each type of SLB and analyzing the inductive properties of the assembled pattern. Note that checking for scalability turned out to be rather simple for the circuits we investigated: this may not necessarily hold for other cases. However, besides scalability analysis, the circuits presented in this work have been tested exhaustively for the indicated input size. Fig. 6d gives the SLBs leading to an ALU, the corresponding circuit is shown in Fig. 6e. The lowest bits of the inputs, *x* and *y*, ($x_0$, $y_0$), form the operator selection ($x_0 y_0$) for the ALU (00 = addition, 01 = XOR, 10 = AND, 11 = OR). The examples presented are taken from a large variety of circuits that were evolved to handle the respective task. It is emphasized that the individual evolved multipliers differ considerably in their self-assembly patterns as well as in the logical functionality of their SLBs. The same holds for the ALUs.

## *2.3  Circuit evaluation*

For a particular run, the size of the board $n = 4n_{bit}$ was held fixed, at a large enough value (e.g. *n = 24, 32 or 48*) to deal with a whole range of different size multiplication problems up to a maximum size. Importantly, and in contrast with other evolutionary approaches, the evolution time (measured in circuit evaluations) proved independent of board size, above a minimum threshold (About four input bits for multiplication, entailing a *16×16*-board.). In this way, we have been able to evolve solutions, tested during the evolution on up to 8×8-bit multiplication (*4×4*-bit usually suffices though), which also scale by the self-assembly process to multiply correctly *16×16-bit*, *32×32-bit* and larger multiplication tasks.

Board-size independence demands for a fitness function also being independent of *n*. Before going into detail, two remarks explaining the underlying ideas are given. Firstly, board-size independence requires that the fitness of a specific circuit not be evaluated exhaustively (by considering all possible $2^{2n}$ inputs), because then the fitness for boards of different size would be calculated with respect to different sets of problems. Instead, the fitness is evaluated with respect to subsets of problems (the test vectors). These test vectors are lists of input pairs and have fixed length $n_{TV}$. Secondly, an *n×n-bit* function has in general a *2n*-bit result string. Taking into consideration all of these *2n* bits would again introduce an n-dependence to the fitness function. The fitness function we devised intrinsically determines the length of a substring (always starting at the least significant, $Z_0$, bit) which then is compared with the corresponding string of the correct result. The actual fitness value of a given circuit is calculated according to the following scheme:

1. Calculate the result-strings ($Z_0$-$Z_{2n}$) for all elements of the test vector and compare them with the correct results.

2. Determine the number $n_{bonus}$ of consecutive bits (starting from $Z_0$), which are correct for all elements of the test vector under consideration.

3. Starting from $Z_0$ and going up to $Z_p$, $p = n_{bonus} + n_{initial\ award} - 1$ evaluate the total number $n_{correct}$ of correctly calculated bits for all $n_{TV}$ input pairs in the test vector. The fitness *f* is then given by

$$f = n_{bonus} + \frac{n_{correct}}{p n_{TV}}$$

For a visualization of this see Fig. 7. This scoring provides a graceful biasing of the n-bit×-bit task towards sub-problem completion and allows progress to be made on large tasks. The evaluation function thus has some features in common with the much studied blocked or "royal road" fitness function[37]. Independently of problem size, individuals have to both connect up external inputs with outputs and compute the appropriate logical mapping (e.g. multiplication). While rewarding the correct completion of all lower bits of a given task first, no problem specific assistance was provided.

## *2.4 Evolutionary dynamics*

To complete our description of genetic self-assembly, the evolutionary process of variation and selection in a population of individuals needs to be specified. Single and independent multiple bit mutation was allowed at different rates for the logical and recognition portions of the genes. In addition, on switching between the four functional categories of logic — function generator, arbitrary MUX, through connection or constantly zero — the logic was smoothed by choosing the closest matching functionality upon mutation. For example, on varying from a MUX to a function generator, the function generator encoding that particular MUX was chosen. This procedure proved reasonably efficient for evolutionary optimization, but is not deemed critical to our success. Secondly we included a variation mechanism involving SLB gene duplication (overwriting an existing SLB gene in the process to conserve sequence length). Thirdly, a general subclass of double mutations in the recognition portion of the genome were chosen at enhanced frequency. These mutations involve twin changes of opposing recognitions bits in juxtaposed edge pairs on two different, randomly chosen SLBs. This structured variation mechanism proved very effective in accelerating evolutionary optimization by inducing frequent changes of the self-assembled pattern.

In consequence, four parameters determine the rate of variation: the bit-normalized mutation rates for recognition and logic, the gene duplication rate and the rate of twin changes of recognition sites. For the multiplier evolution we usually also restricted mutations in the multiplexer bits which activate the function generators, so that an SLB had at most two function generators in use at a time (see Sec. 2.1. above). This was not necessary but also sped up the evolutionary process. In the ALU example this restriction was not employed.

In order to deal with the problem of exponentially increasing test vector sets as the bit length of multiplicands increases, without being restricted to a constant subset, we let the test vectors co-evolve with the circuits. The variation rate for test vectors, the population size and the number of test vectors are then the remaining parameters characterizing the simulation.

# 3 Results

## *3.1 Inductive Generalization*

The main result presented in the paper is the proof that the use of genetically encoded self-assembling components enables the evolutionary design of scalable circuits from examples of correct functionality by inductive generalization. In our simulations, scalable circuits were evolved without information about the structure of the task, using functionally unconstrained logic building blocks, in less than 24 hrs on a PC. Only 16 test tasks and 6 genes for SLBs were required per individual (in a population of 32 individuals), for the circuit to evolve the general ability to solve the posed problem, independently of problem instance size. The complexity of the logic employed in the different evolved circuits varies significantly, but self-assembly seems to provide a natural bias towards more regular logic arrangements, matching our intuition about simplicity. We have not had to introduce any evolutionary constraints favoring minimal or simple circuits.

To get more insight into this phenomenon, we analyze in this section in detail the case of multiplication. Several types of circuit construction problems for multipliers do become formally hard (in NP), for minimal circuit resources[38], and no scalable solutions are then expected. However, without any additional requirements, multiplication has hitherto become increasingly difficult to evolve in larger circuits, because correct solutions are lost in a large search space. We employed a fitness function depending cooperatively on predictions of individual bits in the test products, one that will work for variable length binary products (see Sec. 2.3). The fitness function does not optimize the circuits for compactness, nor does it provide any problem-specific assistance. Instead, we have taken pains to establish a generic unbiased set of logical primitives in order to demonstrate *de novo* evolution of the desired functionality.

Fig. 8 describes the process of inductive generalization for a scalable multiplier circuit. In previous work on smaller multipliers, most optimization time is spent fitting the (drifting) last outstanding multiplications into the almost perfect circuit. By contrast, with genetic self-assembly, there is a time point at which the system captures general features of the ability to multiply (see Fig. 8a). This is accompanied by a sudden increase in the number of completely correctly predicted product bits. Statistical analysis of this phenomenon revealed a clear peak, the "inductive hill" (see Fig. 9a), in waiting times for successive perfection on bits 6 to 8 of the products, with waiting times decreasing to zero for higher bits. The zero waiting time for higher bits reflects the fact that if the genome encodes components which self-assemble to a circuit that solves correctly the lowest eight bits, the system has mastered the task of multiplication in all generality. No time is needed for the evolutionary solution of higher bits, because the circuit is scalable, indicating successful inductive generalization.

Fig. 9b shows the phenomenon of the inductive pass by giving a statistical measure for the waiting time needed for completing multiplication up to a number of $s$ bits in dependence on the mutation rate $r_{TV}$ for the test vector portion of the genome. These waiting times were derived from time series such as in Fig. 8a, giving the maximal fitness in the population. In order to provide a statistically significant picture of the phenomenon of inductive generalization, we defined the time $t_s$ to be the first time at which an individual in the population calculated correctly all the lowest $s$ bits of the problem instances posed by its opponent in a tournament. The waiting times are then given by the difference between $t_s$ and $t_{s-1}$. Two further peculiarities of Fig. 9b have to be noted. Firstly, the runs we investigated were of limited length (20 million individual tournaments) and not all of them yielded a solution. In that case, the incompleted waiting times were set equal to the total length of the run, which explains the plateau for very high and very low $r_{TV}$. Secondly, the data in Fig. 9 represent the third quartile for the corresponding waiting times; this statistic has been derived from 40 simulation runs for each value of $r_{TV}$.

As already stated, these waiting times are defined with respect to time series as shown in Fig. 8. This means that they refer only to those few multiplications coded in the employed test vector, which may appear to be only a weak indication for having solved the problem of multiplication completely. However, the probability of randomly predicting n products for multiplicands of bit length *r* approaches $2^{-2nr}$ for large *r*, already vanishingly small for the number of samples during simulation for *n=4* and *r=6*. Initial concerns about fluctuations making evolution difficult turned out to be allayed by the population dynamics of the test vectors, which served to significantly dampen fluctuations in the evaluation process. Note further that the correctness of the presented circuits in this work is not only justified by the above probability argument; it has been tested exhaustively and the scalability of the circuits has been shown by an analysis of their internal logic. Similar results can be seen for the evolutionary design of ALUs or other scalable circuits, such as adders or binary-to-Gray code converters.

### *3.2 Evolutionary Dynamics*

To investigate the surprising potential of genetic self-assembly further, we traced the time course of the test vectors, multiplication success and structural self-assembly in Fig. 8b and 8c. That multiplication can be performed recursively is well known, but this has hitherto proved difficult to detect from examples. In binary form, the induction may be expressed in terms of the initial conditions a.) and b.) and recursive relationship c.) as in equation A below:

**A. Standard recursive multiplication:**

**a.** )  *a * 0 := 0*

**b.)**  *a * 1 := a*

**c.)**  *a * b := (a * (b mod 2))    +    (2a * [b/2])*

**B. Recursive no carry multiplication:**

**a.** )  *a $*_{nc}$ 0 := 0*

**b.** )  *a $*_{nc}$ 1 := a*

**c.** ) $a *_{nc} b := (a * (b \bmod 2))$ XOR $(2a *_{nc} [b/2])$

where * is the multiplication operation, := represents a definition and the square brackets indicate the "integer part of". In c), the first term corresponds to either definition a.) or b.), since *(b mod 2)* is 0 or 1. The second term involves the product of two terms obtained by multiplication and integer division by 2. These operations are binary shift operations. Although the first term gets larger, the recursion still terminates, because the second term eventually decreases to either 1 or 0. Of the operations, the only operation non-local in the binary representation is the addition operation in c): it involves carry propagation in general. In order to further dissect the inductive principle, which self-assembly enables our evolutionary process to discover, we investigated a subclass of multiplications in which a purely local processing of information suffices. The exclusive or (XOR) operation captures the essence of addition without carry, and hence multiplication without carry as shown in B above. The classification used in Fig. 8 corresponds directly to a dissection of this induction. Multiplications by zero are collected in class I, by unity in class II and by powers of two in class III. Multiplication pairs (a, b), for which the calculation above is the same on replacing + by bitwise XOR, do not require carry-operations and are in class IV. These classes can be understood in terms of circuit sophistication: in order to solve problems in class I, a uniform (zero) output is sufficient, class II needs transfer of the input over the circuit to the outputs, class III requires shift operations, class IV a local form of addition and finally class V involved carry logic. The evolutionary process discovers the inductive principle for multiplying vectors in the simpler classes I-III, then IV, and finally the non-local class V. These classes are first solved with circuits that only work for small examples and then in full generality.

We have shown that the potential of genetically encoded self-assembly for inductive generalization can be exploited by evaluating the fitness of a circuit on a very limited number of mutually exchanged test problems in each step. This procedure, besides being computationally efficient, leads to a co-evolutionary coupling between circuit designs and test vectors (see Fig. 8c) driving the test vector population towards logically convoluted problems. The fact that such a coupling can be observed is no surprise because a genome confronting its adversary in the tournament selection process with a "difficult" problem is likely to have an advantage. The question arises whether this coupling is only an artifact or whether the problem exchange procedure fosters the evolution of structured designs. Fig. 9 and Fig. 10 indicate the latter to be true by presenting statistics for the waiting time for inductive generalization depending on the mutation rate of the test vectors $r_{TV}$. For each value of $r_{TV}$, at least 30 individual runs were performed. For reasons of CPU-time cost and in contrast to Fig. 8, the statistics refers to 6×6-bit multipliers instead of 8×8-bit or larger multipliers. However, several random samples were tested to scale up to 8 bits and no exception to scalability was encountered. One observes that if the co-evolutionary dynamics is disrupted by too large value of $r_{TV}$, the waiting times increase strongly. This behavior is also found for very low $r_{TV}$ resulting in the "inductive pass" observed in Fig. 9b.

In order to understand this, one has to consider that there are two basic strategies for coping with the problem instances in the test vector population. One is to find structured solutions (they may be partial and only be valid for subclasses of the complete set of problems) and the other is to adapt to the actual members of the problem population. The latter strategy may lead to a fast increase of fitness at an early stage or fast adaptation to occasionally emerging new problem instances, but is more vulnerable to fluctuations in the test vector population. This means that if $r_{TV}$ is too low, the evolution of structured solutions is hindered by the relative success of special-case solutions.

This interpretation is corroborated by the results shown in Fig. 11, which basically represents the statistics of the ratio *C* of relative success in classes I-IV over that in class V. This relative success is calculated by counting the number of correct bits divided by the total number of result bits for all *m × m-bit* multiplications, this with respect to the corresponding set of classes (We emphasize that the ratio *C* is not taken only for those problem instances in the test vector population but for all possible *6×6-bit* problems). The number *C* varies of course over an evolutionary run; in order to get statistics over several runs, we took the median value of each individual run and depicted the resulting data set for around 30 runs for each value of $r_{TV}$ in Fig. 11 using box plots. The basic idea is that in case of sole adaptation to specific multiplication tasks, this ratio is expected to be close to one (single instance multiplication is, with the exception of multiplication by zero, of approximately equal difficulty for all bit strings), whereas the evolution of generalizing circuits (with earlier success for classes I-IV than V) is expected to yield a *C* significantly above one. This can in fact be observed in Fig. 11. The median is used as a statistic, because for the time course of evolutionary processes, because it provides better characterizations than mean values as expected for exponentially distributed waiting times in innovative processes[39].

We conclude that a value of $r_{TV}$ small enough to allow the establishment of co-evolutionary coupling of circuit designs and test vectors but sufficiently large to devaluate adaptation to specific problem instances supports the efficient evolution of inductive generalization in self-assembling structures.

## 4  Discussion

Self-assembly encoded structures yield a rather general but biased sampling of possible functions. This was evidenced by the ability of genetic encoding to solve a range of problems, including finding a scalable ALU. In order to distinguish self-assembly genetic guidance from effects due to component genetic encoding, we investigated several different encoding schemes for components. Recognition patterns must be sufficiently diverse to provide a rich set of self-assembly patterns and the number of connections between SLBs must be large enough to allow efficient routing. Encodings in which routing connections (wires) had to be realized as special cases of multi-input logic functions increased evolution time significantly. Restricting the maximal number of combinatorial functions per SLB from four to two proved to be beneficial in the case of multiplier evolution but not for ALUs; but both profited from coding the functions to give a bias for direct I/O-connections (routing). This is achieved via in built genetic biasing, see Sec. 2 and Fig. 3; e.g. simple transfer of an input signal need not be realized by a four input function generator, the evolvable multiplexer allows direct routing. Similarly, the variation mechanism has a strong potential influence on the sampling of new structures. Besides single bit (point) mutations, the addition of duplication events on the genes for individual SLBs was evaluated. This was seen to foster smooth differentiation of component recognition for the assembly process and similarly diversification of logical functionality from simple routing connections. Mutation also change the sampling of new components. Whereas the genetic self-assembly process was relatively robust towards changes in mutation rates for circuit components, the mutation rate of test vectors showed a distinct optimal value (see Fig. 9 and 10). The optimal size of the test vector also turned out to be small (16 problem instances) and to yield complete multipliers within available computation time only for sizes between 4 to 64. Finally, the choice of a modular and scalable fitness function giving a bias towards perfection on subtasks turned out to be important; the fitness function reported proved applicable for all problems investigated.

The simple self-assembly process employed in this paper is a mere caricature of complex regulated physical self-assembly. In particular, there is an additional redundancy and robustness required in physical self-assembly, which is an error prone process. The simplification adopted here, in which self-assembly is only allowed to proceed if both neighbors are already in place, and in which exact matching (taking promiscuous symbols into account) is required, makes the self-assembly algorithm deterministic. In fact, in the present form, the two dimensional build up is formally equivalent to the time course of a one dimensional cellular automata rule (CA), in which the next state is dictated by the two neighboring cells on the previous diagonal. If we restrict attention to the self-assembly of the recognition patterns, ignoring differences of content, the number of such rules can be readily calculated. Ignoring promiscuity symbols, for a binary pattern length $p=2$ this number is equal to the number of possible exposed recognition patterns at an assembly site ($2^{2p}$) raised to the power of the number of possible input patterns on the two binding edges of an assembling block (also $2^{2p}$). The result is $4^4=64$ for p=1 and $16^{16} \approx 2 \cdot 10^{19}$, demonstrating the rapid rise in self-assembly variation as building block edge diversity increases to typical nucleotide levels. For *p=2*, the restricted number *s* of SLBs in the genome provides a stronger limitation for *s<4*. Even without the separate one dimensional self-assembly process for the input layers, the diversity of structures, both periodic and aperiodic, which can be encoded by this self-assembly process is large and include the fractal structures found in the study of CA[40]. The promiscuity bits provide an additional level of fine control over the self-assembly.

However, self-assembling logic is markedly different from CA in its functionality, even in the case where the structural build up of the circuit can be emulated by a CA rule. The self-assembled structures process digital data according to circuit logic rules, whereby routing that extends over several blocks does not rely on extra structures but is realized by the circuit logic of juxtaposed SLBs in the self-assembled pattern. Investigations with non-deterministic self-assembly, in which either error prone assembly (substituting the default block for the correct matching partner with a probability of 1%) or ambiguous matching is allowed (replacing genome-order based choice by a random choice in the case of multiple matching SLBs) demonstrated that the evolution of multiplier circuits can occur in the presence of more realistic self-assembly. This is encouraging for nanoscale physical implementations.

Molecular self-assembly is ubiquitous in living systems, it can give rise to both periodic structures such as multimeric enzymes, microfilaments or viral coats and essentially aperiodic structures such as the ribosome. Self-assembly of linearly connected components occurs in the folding of RNA[41] and proteins. Self-assembly of separate components as in the ribosome[42] is not prohibitively costly for several reasons: modular structure formation is more reliable and self-assembly removes certain intramolecular folding constraints. Employing modular structures leads to large savings in the amount of information which has to be encoded and evolved genetically. Self-assembly may be viewed as one mechanism for generating structure from components. The linear sequence of SLBs in our genome is only of secondary importance in the self-assembly process. Apart from that, at the molecular level, structure generating models play a major role in the biological modeling of development (based on genetic information). The advantages of biomimetic development have been advocated in the area of evolvable hardware[43]. Formal systems for development, such as Lindenmayer systems[44], operate on strings of tokens with parallel replacement rules (operating unlike CAs at the level of substrings rather than symbols) allowing variable length structures to emerge. Self-assembly represents a rather different paradigm, in general involving pattern matching between components, and one which we have shown here to support evolutionary optimization well.

The logical primitives employed in this work are similar to those used in current Field Programmable Gate Array technology, although there is usually a heterogeneous treatment of programmable routing and logical resources on FPGA chips. The major difference lies in the self-assembly process, which would allow a very different production technology from planar lithography, and indeed true three-dimensional structuring. We are conscious that the current article, being based on very idealized components, does not deal with the practicalities of real nanoscale self-assembly. For the connection with current FPGA technology, the self-assembly process can be regarded as a particular type of condensed genetic encoding of circuits which is particularly suited to the task of regular logic generalization. The resulting evolved circuits can be directly mapped to silicon following virtual self-assembly, rather than being self-assembled on-site. Furthermore, current global FPGA rewriting interfaces provide a data bottleneck for reconfiguration, so that the self-assembly algorithm could be used to completely structure FPGA chips with a local reconfiguration mechanism and one involving much less external information. This could greatly enhance the rate at which FPGA configurations could be interchanged in dynamical custom processing applications.

Autonomous logical design has proved hard both for evolution and machines, and has been treated as a sovereign territory for human engineers independently of whether the circuits designed solve NP-hard problems or not. If it were indeed not hard to autonomously design functional systems with given properties (such as circuits), then this would be the method of choice in the electronic design industry. Indeed, evolutionary techniques are not used routinely to design logical circuits, although they have been used for instance to optimally place and route given logic (which incidentally is known to be a NP-hard problem). The appropriate sequence of problems for an analysis of formal complexity involves general tasks with increasingly complex inductive principles. Current solutions to even simple families of such problems have only been rationally designed by humans and not solved autonomously. Finding inductively scalable solutions for circuits from examples has been an unsolved problem to date. If the underlying problem was also NP-hard we would not even expect scalable solutions to exist. In fact, certain types of multiplication problems have been classified as NP-hard and others not (in particular those with minimal resource utilization are typically hard). The family of multiplication problems we addressed does have an inductive solution and is not hard in this sense. Finding such a solution autonomously from examples, from amongst all possible logical mappings, as our evolving system does, is indeed a hard problem. Genetic self-assembly also manages to discriminate between special and general solutions to the problem in the presence of limited information about correct multiplication.

Finally, we are all aware of the difficulty children have in learning to multiply. In fact many children can multiply the examples learnt by heart long before they are able to multiply larger numbers. This is an example of a specific non-scalable solution to learning the multiplication problem (often stopping at *12\*12*). Only then, they learn an algorithm for multiplication by paper and pencil. Considering this, it is noteworthy that inductive generalization is achieved at the same level of problem sophistication, namely after perfection of the 6-8 lower result bits (see Fig. 9).

## 5   Conclusions and Outlook

Genetic self-assembly facilitates the evolution of inductive generalization by providing a compact encoding of components for scalable structures, by providing a coding bias towards a diverse family of symmetric structures, both periodic and fractal (thereby mapping internal logical structures of a problem into geometrical patterns), by allowing neutral evolution and gene duplication via unemployed components, by providing a natural pressure towards reuse, by allowing a meaningful compromise between being able to encode every possible topology and having sufficient flexibility to find an appropriate one, and finally by allowing a clear separation between genetic information for functionality and for assembly.

The results of this paper give rise to the conjecture that the widespread appearance of self-assembly in biology, besides the advantages of reusability of components for different purposes, is also beneficial for efficient evolvability. This aspect is not only of relevance for the understanding of existing biological systems, but may find applications in recent attempts towards artificial cells.

Good logical design appears to be poised between order and disorder, between regular scalable structures and specific "glue logic". This article confirms that it makes sense to make use of natural computational principles, which share this tension, such as self-assembly, if we wish to harness the power of evolution in the design process. Human generalization and discovery appears immensely more powerful, but the authors hope that this work will point towards a further series of more sophisticated self-assembly processes which can help bridge the gap in our understanding how evolution has achieved the marvelous functionalities observed in living systems.

**Acknowledgements** We wish to thank M. Eigen and the Max Planck Gesellschaft for supporting our first enquiries in evolving logic. RF was supported by the Swiss National Science Foundation and JSM by the Deutsche Forschungsgemeinschaft during the early stages of this development. The European Unions FP6 IST-FET Integrated Project PACE provided financial support in the latter stages of the work, in particular for utilizing self-assembly and evolution for information processing. We also wish to thank M. Bedau, N. Packard, and S. Rasmussen for discussions during a sponsored visit at the Santa Fe Institute, New Mexico, and R. Walker, G. von Kiedrowski and C. von

# Appendix

In this appendix, we present technical details of our implementation. They proved to lead to efficient evolution, but as long as the general features discussed in the main text are preserved, alternatives may well be equivalent.

## *Logic functions of SLBs*

In principal, it would be possible to simply equip each output o0-o3 of an SLB (Fig. 3a) with an evolvable four-bit function generator. However, in our implementation, the genetic encoding of internal logic of each block was chosen both to permit arbitrary combinatorial functions of the four inputs at each of the four outputs and to give significant probabilities to increasingly simple sub-classes of mappings: arbitrary direct connections of inputs to outputs (MUXs), straight through connections and constant output values. This was achieved by, for each output, encoding a function as given in Fig. 3b. A complete four-bit function generator is only one possibility; whether this or a simpler mapping is selected depends on the evolvable configuration of two multiplexers.

## *Recognition sites*

Local recognition and consequent adhesion is modeled in a "soft" variant, allowing different qualities of match, see Fig. 3c. Each edge of a SLB has two recognition sites, that can either host up to two plugs (right and lower edges) or two sockets (left and upper edges). Plugs and sockets each come in two types (encoded by one bit), whereby there are two different possible plug/socket pairs. If there is no plug/socket at a specific position (encoded by a second bit), this position becomes promiscuous. A connection is prohibited if opposing plug/socket pairs do not match and each exact (non promiscuous) match is given a constant binding energy (or quality contribution), which will be of relevance for resolving ambiguities in the self-assembly process. The genetic encoding then involves *4\*(2+2)=16* recognition bits (for the four sides) and *4\*(16+2+2)=80* logic bits (for the four outputs) giving a total of 96 bits per SLB for recognition and logic together. The number of recognition bits is explained by the fact that each of the four recognition sites may or may not contain plugs/sockets of two different types at two different positions. The logic bits comprise a four-bit function generator (16bits) and two four-input multiplexers (twice two bits) for each of the four outputs.

## *Edge SLBs*

The self-assembly of the edges, extending evolutionary flexibility, follows equivalent rules and uses the same sort of recognition sites as those of the board. The functionality of edge-SLBs is given by two multiplexers that either transmit the two input signals or connect them to ground (Fig. 5a). The structure of the board (Fig. 5b) is slightly altered; recognition sites in the upper left corner initiate the self-assembly process and the inputs previously connected to ground (Fig. 4a) are now also used for connection to the environment. Edge-SLBs self-assemble according to analogous rules as for the bulk-SLBs (Fig. 5b). Typically, when self-assembling edges were employed, 2 to 6 such edge-SLBs were included in the genome, each requiring *3\*(2+2) + 2 = 14* bits (edge blocks have only three self-assembling edges). Edge-self-assembly turned out to be necessary for the evolution of general ALUs on the available timescale. In summary, self-assembly is initiated at a (zero-dimensional) corner point, proceeds by installing one-dimensional rims and then passes on to two-dimensional assembly of the circuit. The order is immaterial as long as one requires left and upper components to be installed before self-assembling components to the right and down. Obviously, this scheme can be extended simply to higher dimensions. Finally, the geometry in Fig. 5 implies that there is a difference between horizontal and vertical edge-SLBs. There is no difference in encoding: vertical and horizontal edge-SLBs are simply mirror-images.

# Tables

**Table 1: Complete list of parameters for genetic-self-assembly.**

| Complete Parameter List | | | |
|---|---|---|---|
| General parameters | *Symbol* | *Default Value* | *Values for ALU* |
| population size | $N$ | 32 | 32 |
| evolution time in generations | $t_G$ | 100000 | 100000 |
| number of SLBs in genome | $n_{SLB}$ | 6 | 10 |
| number of test instances in genome | $n_{TV}$ | 16 | 32 |
| length of recognition pattern on edge | $l_{rec}$ | 2 | 2 |
| size of problem in argument bits | $n_{bit}$ | 8 | 8 |
| Mutation | | | |
| probability of mutation in SLBs | $R$ | 0.7 | 0.7 |
|     probability of mutation in function | $r_{func}$ | 0.5 | 0.3 |
|     probability of gene doubling | $r_{GD}$ | 0.05 | 0.05 |
|     probability of match mutation | $r_{PS}$ | 0.05 | 0.05 |
|     probability of mutation in recognition | $r_{rec}$ | 0.4 | 0.25 |
| probability of mutation in test vector | $r_{TV}$ | 0.01 | 0.01 |
| *Fitness function* | | | |
| fitness bit look ahead | $n_{initial\ award}$ | 2 | 2 |
| *Edge SLBs (Optional)* | | | |
| number of edge SLBs | $n_{eSLB}$ | 0 | 3 |

| probability of mutation in edge mask | $r_{efunc}$ | 0 | 0.01 |
| probability of mutation in edge recognition | $r_{erec}$ | 0 | 0.24 |

The table explains the parameters and lists the default values used in Fig. 6

**General Structure**: A population of $N$ individuals is observed over $t_G$ generations. The genome of each individual codes for $n_{SLB}$ different SLBs plus the default block. Maximally $n_{FG}$ out of 4 outputs are connected to four input function generators. The number of positions in the recognition sequence is given by the length $l_{rec}$ per edge. Note the self-assembly array size of $s_{board}$ * $s_{board}$ is arbitrary when scalable solutions can be found; $s_{board}$ is a multiple of four and yields a board which hosts a $n_{bit} \times n_{bit}$ – multiplier, whereby $n_{bit} = s_{board}/4$. We used *24x24* or *32x32* boards representing *6 × 6-bit* or *8 × 8-bit* multipliers respectively. The computation time is increasing at most quadratically with $s_{board}$.

**Mutations** on replicated individuals were introduced by repeating indefinitely until failure (failure with probability $R$). A mutation event was structured as follows:

Four possible types of mutations: a single bit mutation in a SLB with prob. **r_func** (each of the 80 logic bits from a randomly selected SLB is chosen with equal probability); a gene duplication with prob. $r_{GD}$; the selection of a plug/socket pair with the switching of one bit on their respective sequence with prob. $r_{PS}$; a change of one of the 16 recognition bits with prob. $r_{rec}$. The system is not very sensitive to changes in these parameters. Note that the mutation probabilities add up to one.

**Test Vectors:** A test vector contains $n_{TV}$ pairs for input. Both numbers are between 0 and $2^{nbit}-1$ for $n_{bit}$ bit multiplication. Here mutation occurs with prob. **r_TV** and results in the random exchange of both numbers of a randomly chosen pair.

**Fitness Function**: The fitness function is determined by one parameter, namely the number of initially awarded bits $n_{initial\ award}$.

**Self-Assembling Edges**: Additional parameters are necessary if one uses self-assembling edges. First of all, $n_{eSLB}$ denotes the number of edge SLBs (plus one default edge SLB). Recognition sites are assumed to be of the same length as for the SLB used for the board. Additional mutation probabilities are $r_{efunc}$ (giving the mutation probability for a masking bit) and $r_{erec}$ (describing changes in the recognition sequence of the edge SLBs. Note that gene duplication and manipulation of plug/socket bits also applies to edge SLBs.

# Figure Captions

**Figure 1. Overview: genetically encoded self-assembling circuits** (the figure refers to a multiplier) a) the self-assembling logical blocks (SLB) are genetically encoded and may be expressed (b) in whatever numbers are necessary for the self-assembly (c) of a complete circuit (d). Each side of an SLB exposes up to two heterophilic adhesion sites of two different specificities. Recognition is achieved by the requirement that, if both are present, two opposing adhesion sites have to be of the same specificity. Each SLB is equipped with four inputs (i0-i3) and four outputs (o0-o3), whereby the outputs are functions of all inputs. Recognition need not to be maximal for all sides of the SLBs; this flexibility turned out to yield a rich and evolutionary efficiently exploitable variety of patterns. The self-assembly process results in a complete feed-forward circuit (d) that gets input from two sides and yields a result at a third. The circuit can have any desired size: the self-assembly process delivers correct multipliers for any extension. For the precise coding of the logical functions, the wiring of the complete circuit to the environment, the details of the self-assembly algorithm (such as how to avoid ambiguities) and a complete list of suitable parameters see the Sec. 2 and Table 1 and Figs 2, 3, 4 and 6. The selection process is shown in (e). In a spatially homogeneous population pair-wise tournament selection is applied, whereby the fitness of each individual is determined by evaluating a rating achieved on the tasks encoded in the opponent's genome. The winner is replicated (in an error-prone process) by overwriting the looser. Microscopic planar self-assembly of electronic components (f) is already possible under suitable conditions.

**Fig. 2. Schematic of basic selection process and genetic encoding of self-assembling logic blocks (SLBs).** Two individuals are chosen from the population at random and their respective genomes are decoded in a three-step process: first the recognition sequences determine the self-assembly process, then the internal logic of the self-assembled blocks defines the functionality of the circuit and finally by evaluation of the circuits on the decoded opponent's test

vector. The individual with greater fitness wins. If two individuals have the same fitness, then a random test vector is chosen and the test repeated. If the values are still the same, one of the two individuals wins at random.

**Figure 3. Self-assembling logical blocks (SLB):** a) the basic structure of a SLB. Four outputs give signals that are logical functions of the four inputs and the edges of a SLB are equipped with recognition sites, which can be imagined as plug/socket structure or heterophilic adhesion sites. b) The evolvable logic of each output favors routing in a generic manner. A multiplexer chooses one out of four signals: one out of the four inputs, the input directly opposed to the output under consideration, a default signal (ground, which represents zero throughout the whole paper), or a fully evolvable four bit function generator. c) Recognition is realized in a soft variant, plugs and sockets have to match, but not every position in the recognition sequence needs to be occupied by a recognition element. The number of matching plug/socket pairs gives a binding energy or quality relevant for resolving ambiguities discussed in Sec. 2.2. d) A default SLB, only transmitting signals, is used if no other components match with a given configuration of recognition sites appearing in the self-assembly process.

**Figure 4. Self-assembly of a circuit.** The board serves as substrate for self-assembly and provides connections to the environment. The outer rims of the board provide the recognition sites (here represented in a simplified form) for initializing the self-assembly process. Shown is only a small board for a *2 × 2-bit* problem. The genome is translated and as many copies of each type of SLB are produced as are required to complete the circuit. A component is added, if the left and upper edge can brought into match with already placed SLBs. Possible matching ambiguities are resolved as described in the text.

**Figure 5. Self-assembling edges.** Self-assembling and evolvable outer rims of the board provide higher flexibility and turned out to be necessary for the evolution of some types of circuits such as e.g. ALUs. a) The edge blocks embody two functionalities, beside the recognition sites for their own self-assembly. First, they can either transmit an input signal from the environment or provide a connection to ground, controlled by an evolvable multiplexer. Second, the recognition sites they present towards the interior of the board allows non-trivial patterns for initializing the self-assembly of bulk SLBs. b) The connection of the board to the environment is expanded, and the initialization of the edge self-assembly is done from the upper left corner. The edge self-assembly process, besides being restricted to one dimension, is equivalent to that of Fig. 4.

**Figure 6. Evolved multiplier.** The figure refers to a multiplier evolved using parameters as given in Table 1. a) All six SLBs plus the default block coded by the genome. Direct connections are given by wires, connection to zero by use of the ground symbol and the function generators are represented by boxes containing four HEX-digits. It is implicitly assumed that the function generators are connected to all four inputs. b) Scheme for translating the 16 possible outcomes of a four-bit function generator into a four HEX-digit number. c) Assembly of the multiplier, finally resulting in the complete scheme given in Fig. 1d. d) The SLBs for an evolved ALU. e) The ALU-pattern.

**Figure 7. Scalable fitness function.** Visualization of the fitness evaluation for a test vector of size 8. The bonus is determined by the number of successive least significant bits calculated correctly for all problem instances in the test vector (filled circles: correct bits, open circles wrong bits). The fitness is then given by the bonus + the relative amount of correct bits for the first (bonus + initial award) bits, in the figure bonus = 4, initial award = 2, fitness = 4 + 37/48. This fitness function is independent of problem size. For all shown simulation, the initial award was chosen to be two.

**Figure 8. Typical time course of population evolution** for the self-assembling 8-bit multiplier circuit shown in Fig. 1. **a)** Time course of maximum fitness in the population: in black evaluated using its own test vector, in gray evaluated using random test vectors. The block fitness function leads to stepwise enhancements in fitness. Note the large jump in fitness at generation 82000, corresponding to discovery of a general solution to the problem. **b)** Population average of success frequencies for individual product bits 0-15, with whiter fields indicating more success, for each of the categories I-V of multiplication tasks (see text). Solutions to these tasks are successively more difficult to evolve. Additionally, the outer bits are logically less convoluted than the inner ones and consequently are correctly multiplied first. **c)** Relative frequencies of multiplications in the test vector attached to the fittest individual in the population grouped in five categories as in b) with whiter fields indicating higher frequencies. At generation 19000, b) indicates a jump in performance in the class I-IV problems. Consequently, the test vectors concentrate on the final residual category V, as shown in c). When these multiplications have been solved, at about generation 82000, the coevolving test vectors begin to repopulate the other categories again.

**Figure 9: Inductive pass: logarithm of first waiting times for transitions between completion levels in the fitness function at various values of the test vector mutation rate $r_{TV}$.** Completion level s corresponds to first complete prediction of all product bits up to *s*-th bit for all the multiplications of an opponent's test vector. Shown are the third quartile values for the first time $t_s$ to level *s* fitness from level *s-1*, means the time three quarters of all runs needed to reach fitness level s, after having previously reached level *s-1*. A zero value for a waiting time corresponds to multiple level improvements. Reproducible generalization ability is reflected by the fact that, after reaching a certain level of circuit evolution, the system needs no further intermediate steps (and therefore time) to find the final complete self-assembling *6bit×6bit* multiplier circuit. This would extend to arbitrarily large multiplier circuits, and therefore traversing the "inductive hill" means implementing the general concept of multiplication. This inductive hill is shown in **a**), where the waiting times for a specific value of $r_{TV} = 0.01$ are shown. Figure **b)** shows the waiting times additionally in dependence on $r_{TV}$, whereby a) corresponds to one slice. The dependence of the height and extension of the inductive hill on $r_{TV}$, gives rise to the term "inductive pass". A run is halted after 625000 generations (equals 20 million individual evaluations in a population of 32), therefore this maximal value in the plot indicates that less than

three quarters of all runs did succeed. All parameters are the ones given in Table 1.

**Figure 10: Box representation of the statistics of the time $t_{compl}$ required for evolutionary circuit completion for various mutation rates of test vectors $r_{TV}$.** Each of the (at least 30) runs is terminated either by success or by reaching 625000 generations (equals 20 million evaluations in a population of 32); the termination means that if more than three quarter of the runs did not succeed, the upper three quartiles are quenched into the upper whisker. With the exception of $r_{TV}$, parameters are given by the values for the multiplier in Table 1, and $n_{bit} = 6$.

**Fig. 11**: **Statistics of success ratio.** Shown is the statistics of the median ratio of bitwise relative success *C* with respect to problems in class I-IV to the relative success in class V. The ratio was calculated by evaluating the number of correctly calculated bits with respect to all possible problem instances for $n_{bit} = 6$, not only employing those present in the test vector population, and, for each individual run, averaged by taking the median over the whole time course, therefore giving one number per run. The settings are the same as for Fig. 8. One observes *C* to be significantly larger than 1.0 for higher $r_{TV}$ indicating the evolution of calculating structures (see text).

# Figures

## Figure 1

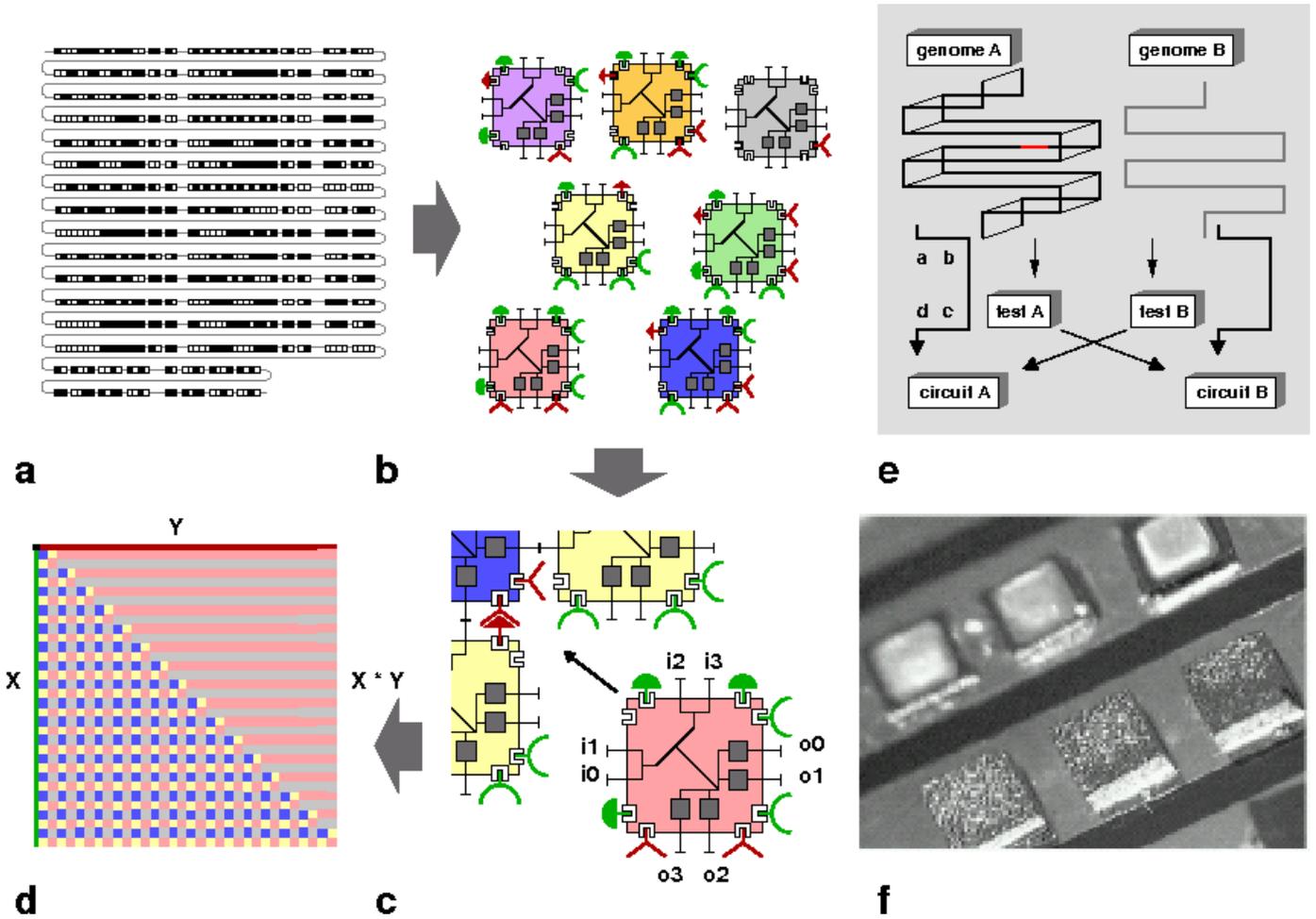

**Figure 2**

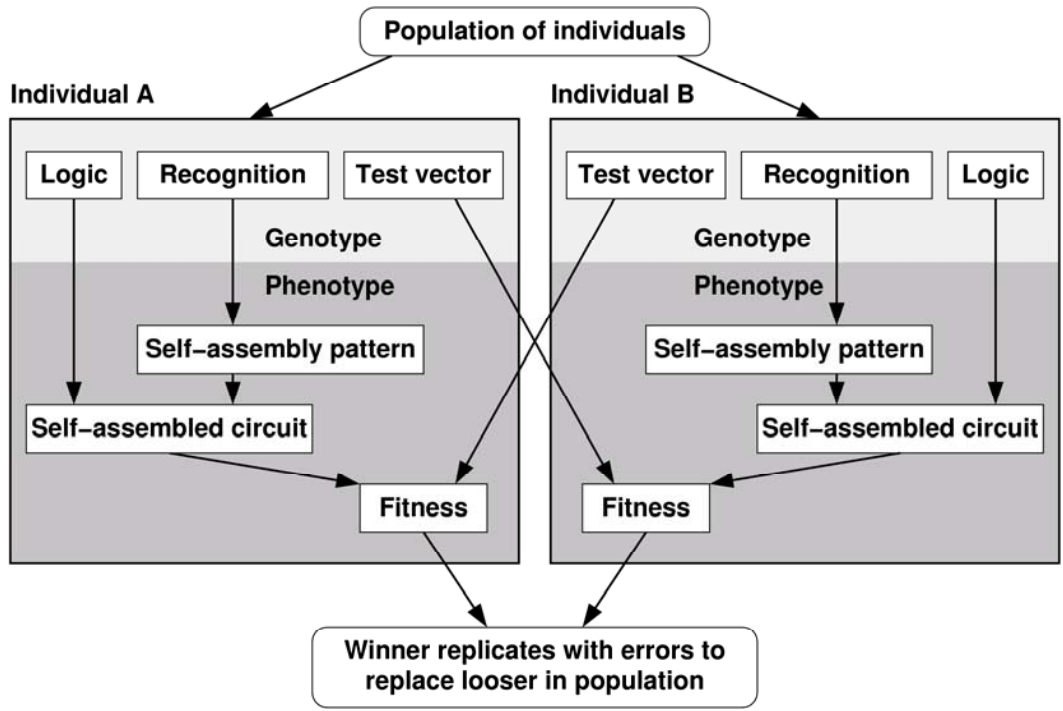

**Figure 3**

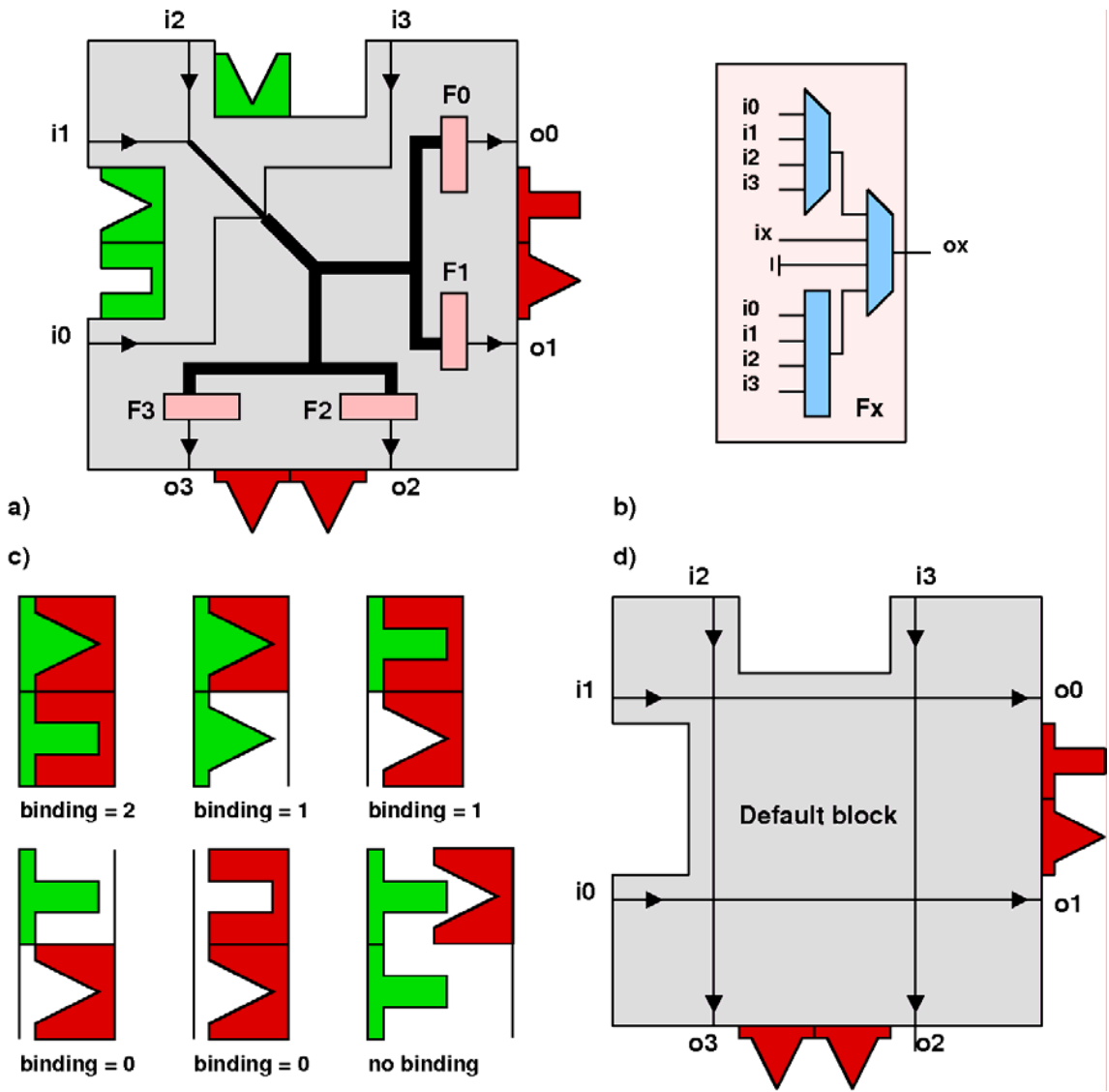

**Figure 4**

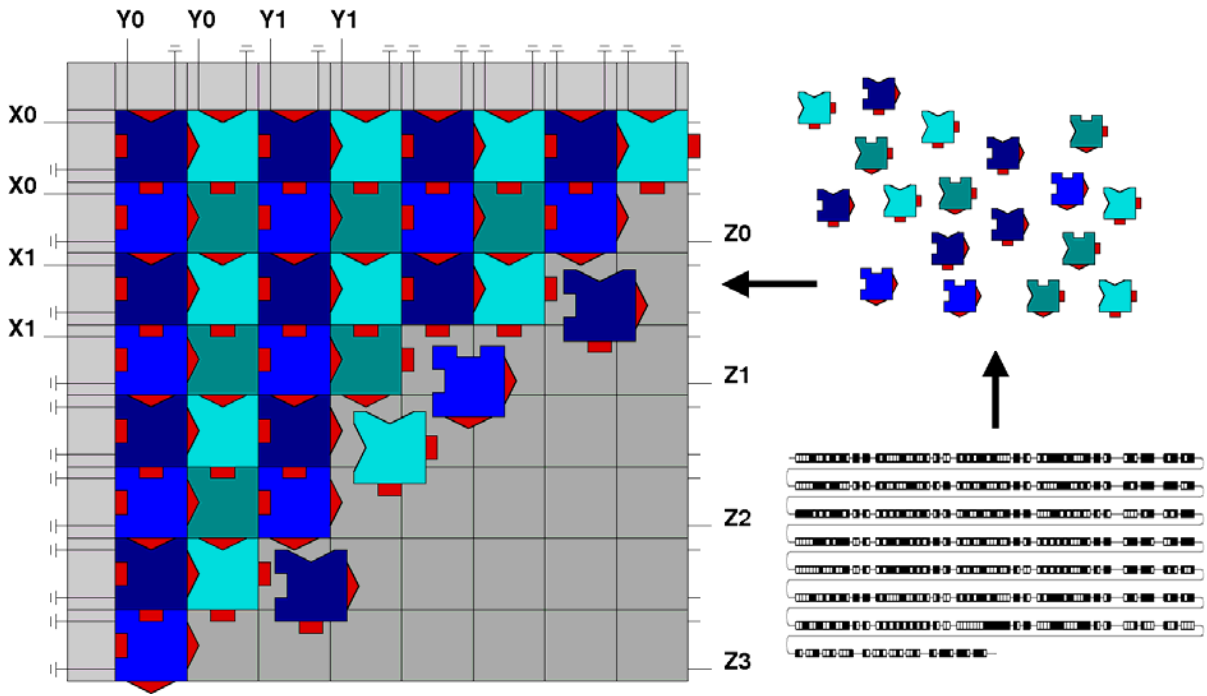

**Figure 5**

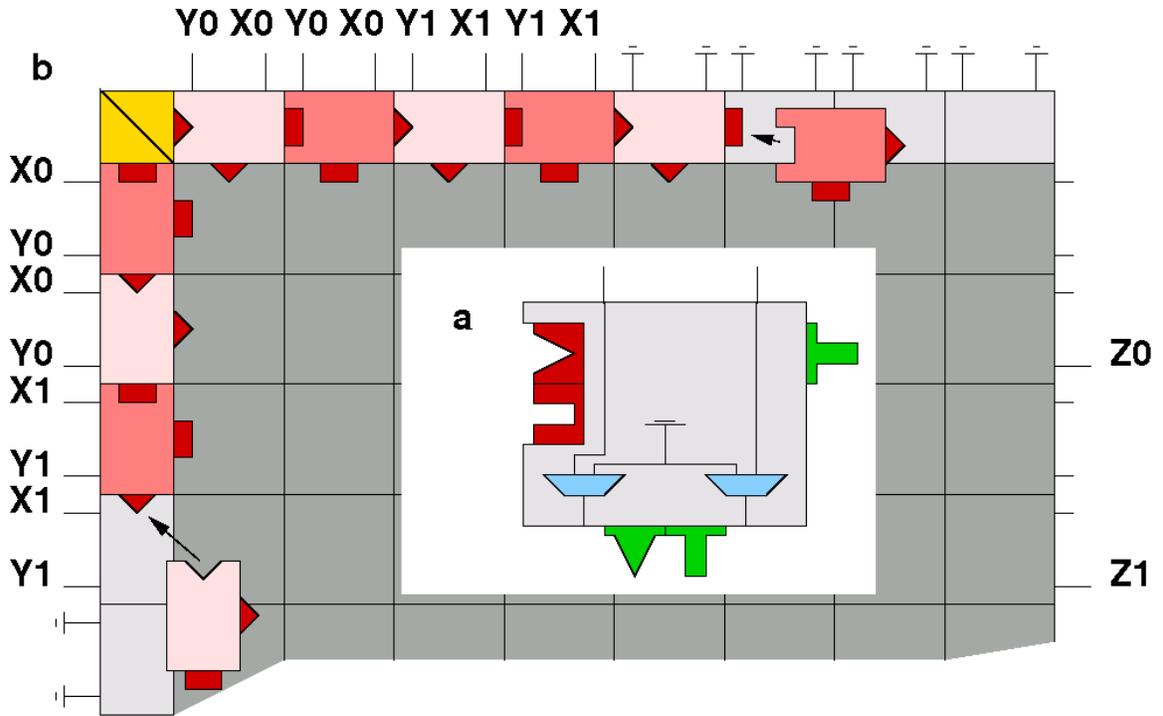

# Figure 6

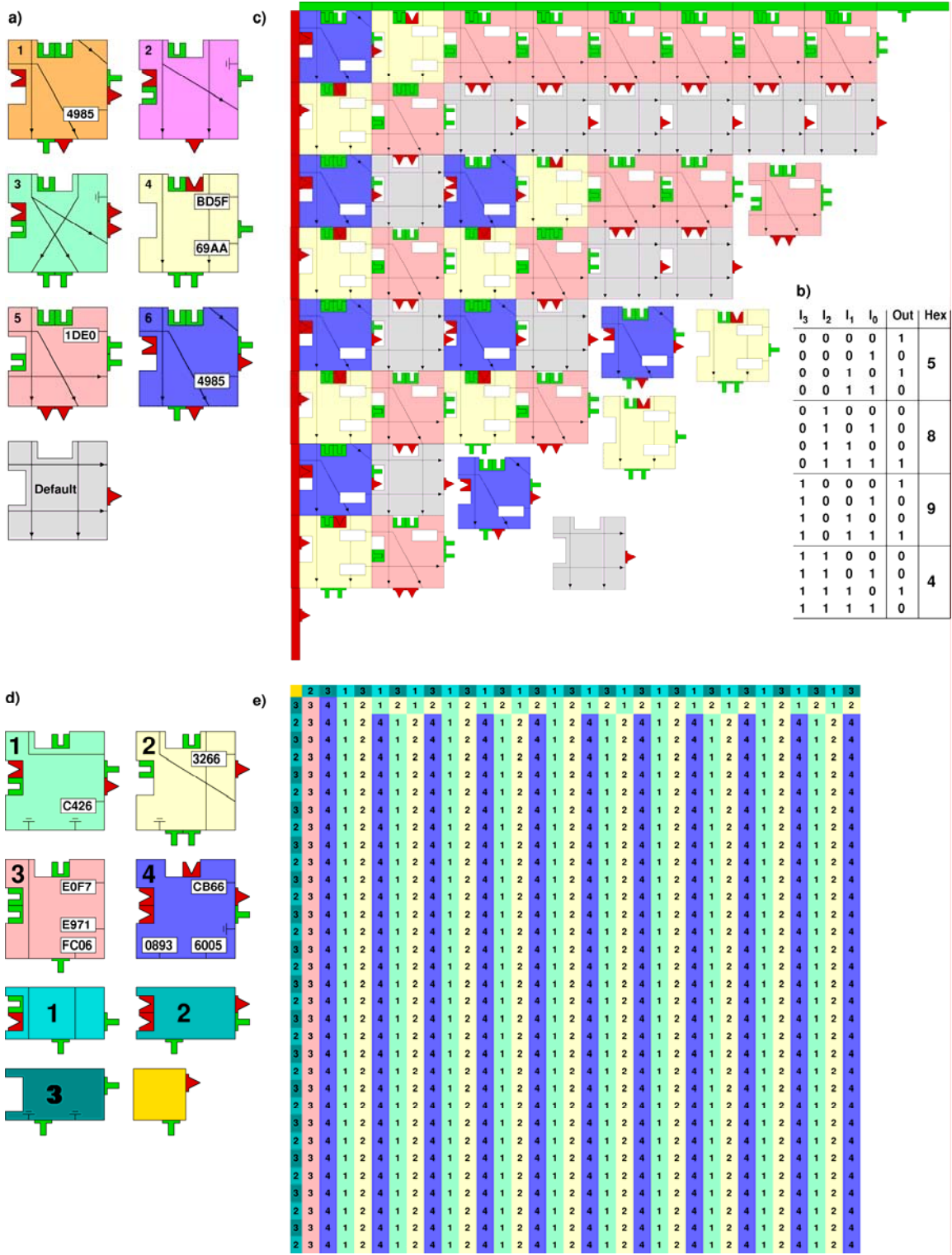

**Figure 7**

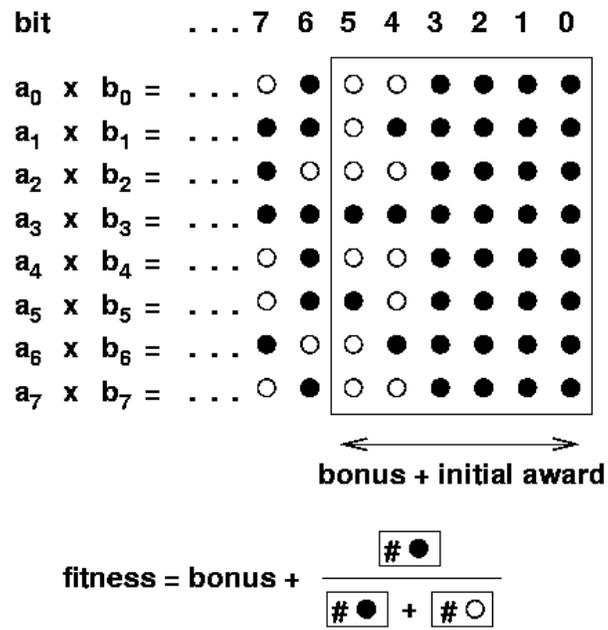

**Figure 8**

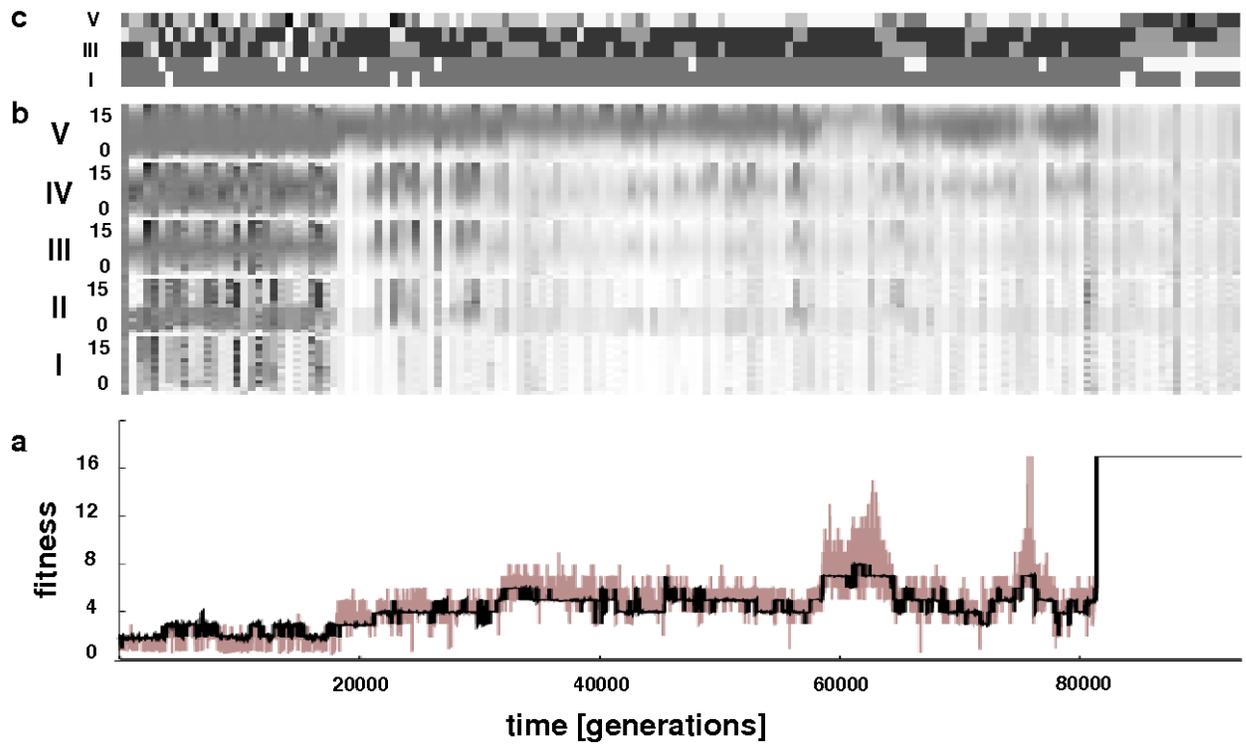

**Figure 9**

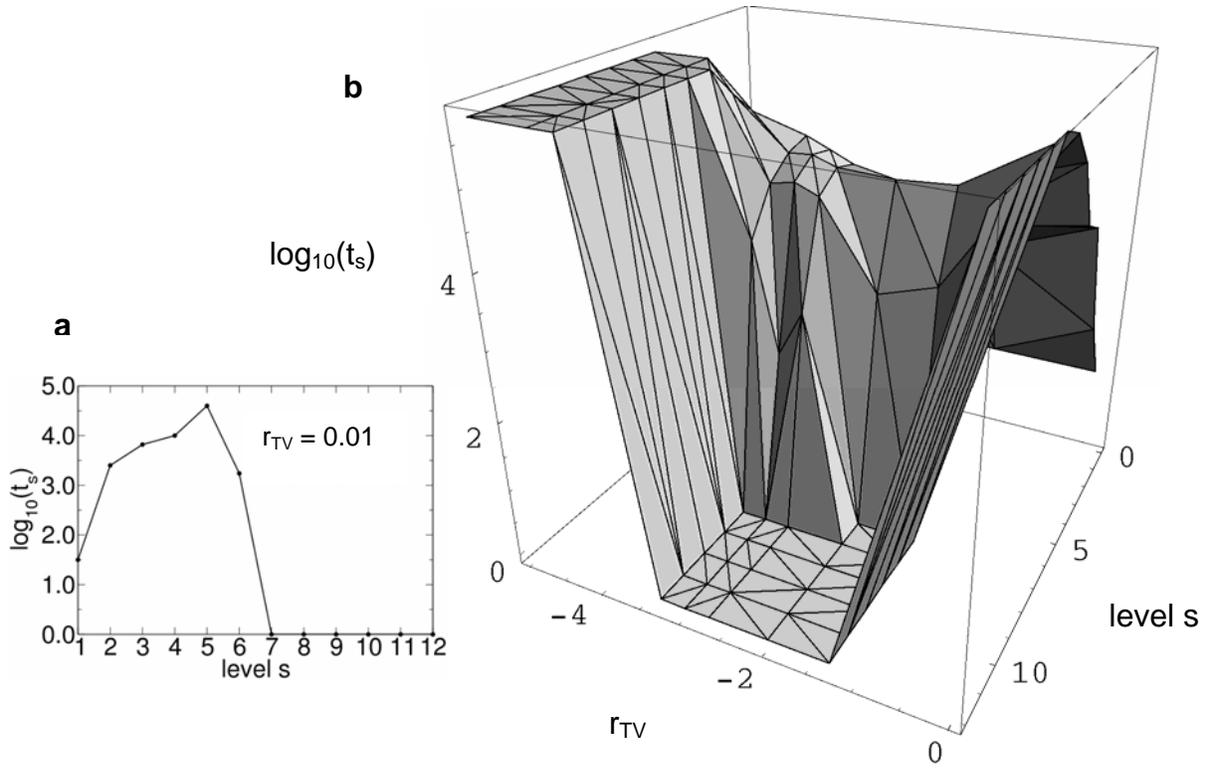

**Figure 10**

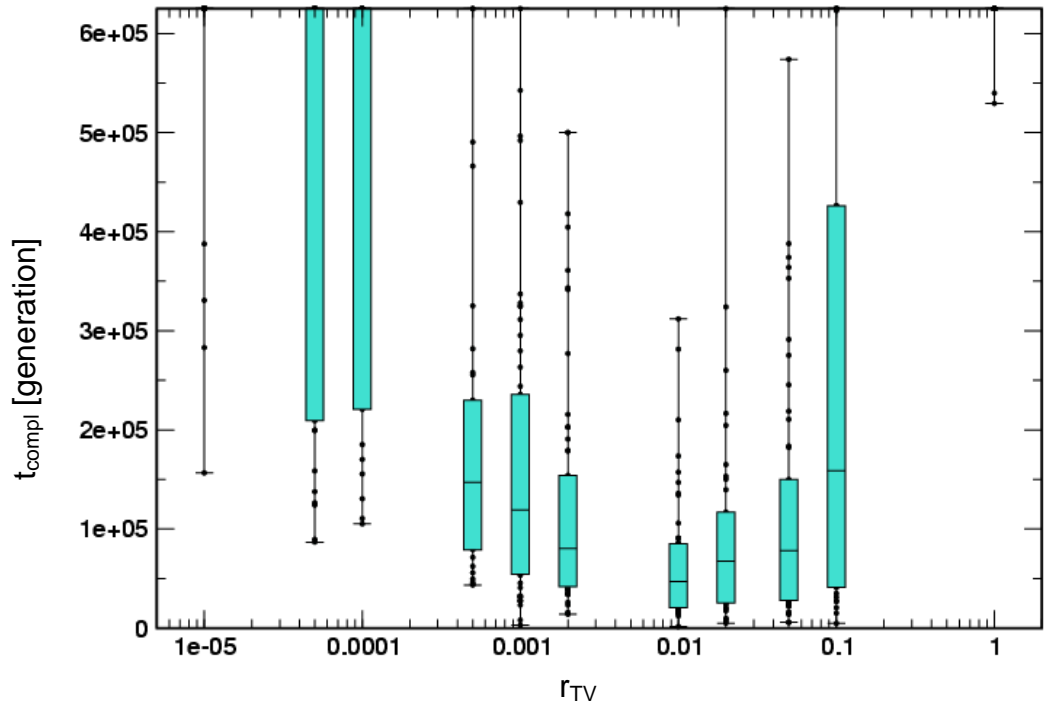

**Figure 11**

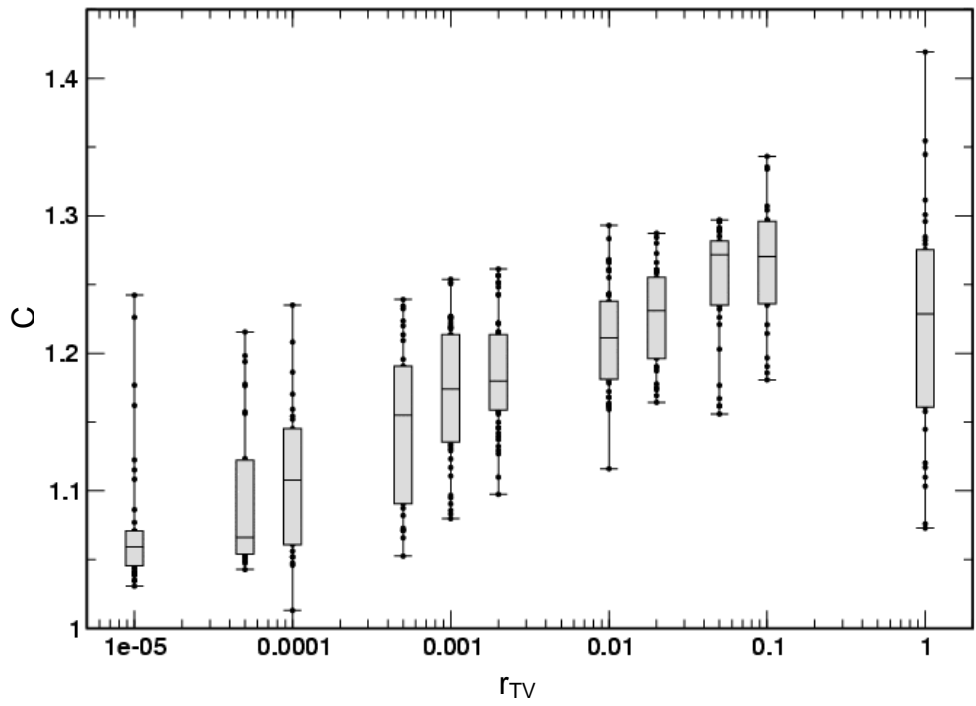